\documentstyle[preprint,epsfig,aps]{revtex}

\begin{document}
\hfill IFT-P.050/97 

\hfill IFUSP/P-1279
\vskip 0.6cm

\centerline{\bf Are High Energy Cosmic Rays Magnetic Monopoles?}

\centerline{C.O. Escobar$^{1,2}$ and R.A. V\'azquez$^{1,3}$}

\centerline{\it $^1$ Instituto de F\'{\i}sica, Universidade de S\~ao Paulo,}
\centerline{\it Caixa Postal 66318 }
\centerline{\it 05315-970 -- S\~ao Paulo, S. P. }
\centerline{\it Brazil}

\centerline{\it $^2$ Instituto de F\'{\i}sica ''Gleb Gataghim'}
\centerline{\it Universidade Estadual de Campinas, Unicamp}
\centerline{\it 13083-970 -- Campinas, S. P. }
\centerline{\it Brazil}

\centerline{\it $^3$ Instituto de F\'{\i}sica Te\'orica,}
\centerline{\it Universidade Estadual Paulista}
\centerline{\it Rua Pamplona, 145}
\centerline{01405-900, S\~ao Paulo, S.P.}
\centerline{\it Brazil}

\begin{abstract}
We argue that magnetic monopoles can not be associated to the highest energy
cosmic rays as recently suggested. Both the observed spectrum and the arrival
direction disagree with observation.
\end{abstract}

\section{Introduction}

The highest energy cosmic rays are subject to several speculations about their
origin and acceleration mechanism. Perhaps the most studied models, besides
conventional models, are the top-down models (TD), where the particles are
generated at extremely large energy and loose part of it by interacting with
the medium. A central characteristic of these models is the (observed) flat
spectrum they predict at the largest energies. The way TD models obtain the
power to accelerate particles depends on the model but in all of them it is
related to some topological defect produced with gauge symmetry breaking of a
Grand Unification Theory in the early stages of the Universe. 

Recently, it was suggested that these cosmic rays could be due to relativistic
magnetic monopoles interacting with the air \cite{Weiler_1,Weiler_2}. Here it
is assumed that it is the monopole which directly will produce the shower by
interacting with air in some not completely specified way. The monopoles could
originate a shower by some catalysis mechanism or initiate a electromagnetic
shower. It is assumed that showers induced by monopoles have the same
characteristics as regular showers. The monopoles easily pick up energy in the
galactic magnetic fields due to the large charge ($\alpha_M \sim 1/\alpha$) of
the monopole. With a relatively low monopole mass M$\sim$ 10$^{10}$ GeV typical
galactic magnetic fields will accelerate the monopole to $\gamma \sim 10$ in a
300 pc distance, easily acquiring enough energy to produce the observed events.
The proposed mechanism is based on three ''coincidences'' \cite{Weiler_2} i) In
300 pc the monopoles will acquire enough energy to produce the showers, as said
above. ii) The observed HECR flux is of the same order of magnitude as the flux
allowed by the Parker limit \cite{Parker_1,Purcell_1} on monopoles for low mass
monopoles, M$\sim$ 10$^{10}$ GeV. iii) The limit on the mass of the monopole
given by the Kibble mechanism \cite{Kibble_1} for their generation in the early
Universe gives the same order of magnitude as the previous points, so that
monopoles do not overclose the Universe.

This interesting suggestion could avoid problems with the GZK cut off which are
present for other scenarios and with the maximum reachable energy. We argue,
however, against this idea. We will see below that monopoles will have a
different spectrum, much flatter, and moreover a visible anisotropy in the
direction of the magnetic fields. 

\section{Simulation}

We have developed a Monte Carlo program to propagate monopoles in the galactic
medium. The galactic magnetic field is parametrized using the models given by 
Stanev \cite{Stanev_1} and Vall\'ee \cite{Vallee_1}. For our calculations we
will used both models given in ref.\cite{Stanev_1}, the ASS\_S and BSS\_A
models. The ASS\_S model (model A, from here on) is axisymmetric and symmetric
in the $z$ component whereas the BSS\_A model (model B) is bisymmetric and has
odd parity in the $z$ component. In our simulations we will neglect the $z$
component. A random magnetic field is added to simulate the random component.
The random field is chosen with a random direction in a lattice of 300 pc and a
magnitude chosen to be of the order of 20 \% of the regular component. 
Monopoles are followed in the magnetic field by numerically solving the
equations of motion using an adaptive Runge Kutta method. The step is forced to
be always much lower than the size of the random component lattice, 300 pc, in
order not to overstep the random lattice.

\subsection{Scenario A}

\begin{figure}
\epsfxsize=10cm
\begin{center}
\mbox{\epsfig{file=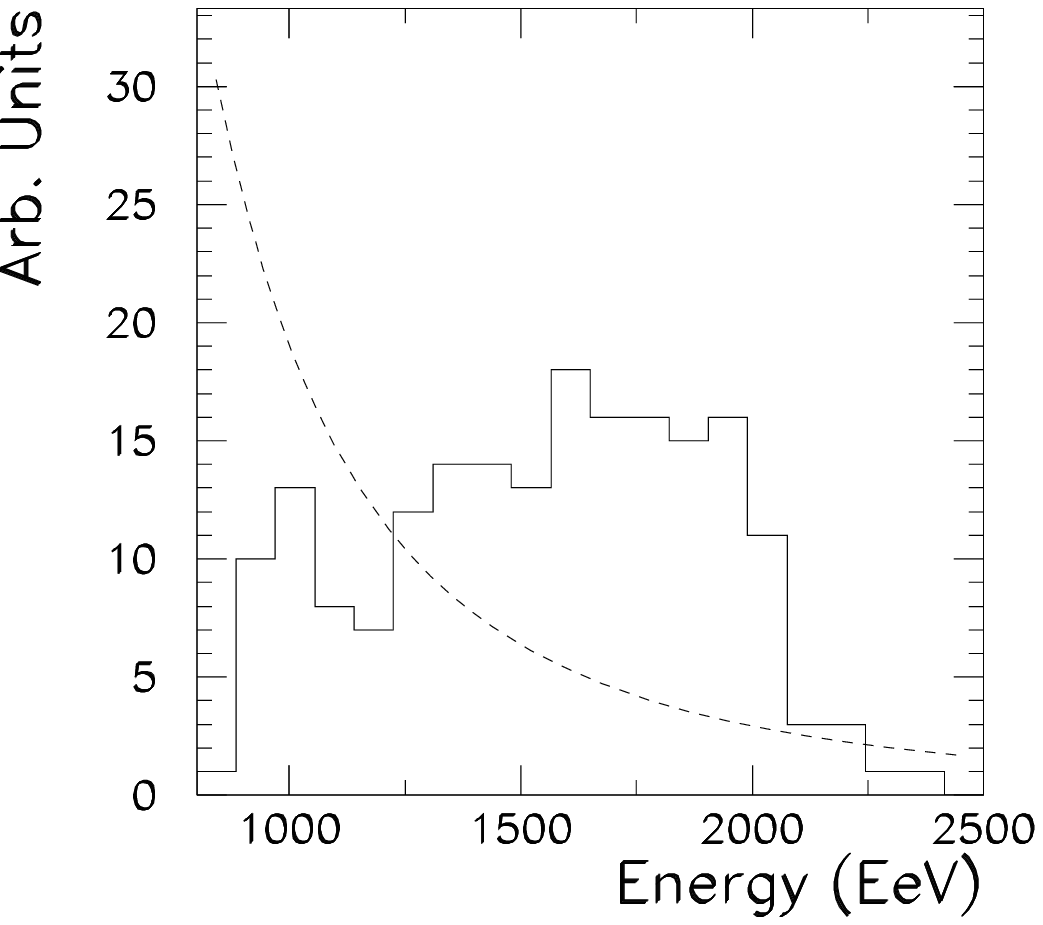}}
\end{center}
\caption{Energy spectrum of monopoles coming from the center of the Galaxy for
the Model A of the magnetic field. Dashed line shows a typical power law
HECR spectrum $\sim E^{-2.7}$, normalized to the same number of events.}
\label{spectrum}
\end{figure}

Let's assume first that near the center of the galaxy there is an outgoing
isotropic flux of monopoles. This would be the case if, for instance, a
supernova exploded liberating its (possibly bounded) monopoles, as pointed in
ref.\cite{Weiler_1}. Monopoles are then accelerated and followed until they
reach a distance equal to the distance to the Earth from the center of the
galaxy. We found that for both magnetic fields the spectrum of the monopoles
reaching Earth is very different from the observed spectrum. In
fig.\ref{spectrum} we can see the spectrum of the hypothetical monopoles for
magnetic fields model A (figures for the model B are similar). For the sake of
comparison we also show a typical cosmic ray flux spectrum (with slope $-2.7$)
and normalized to the same number of events. We see that for monopoles the
spectrum peaks at a very high energy, indeed much higher than the maximum
observed energy. In order to see 8 events with energy higher than 100 EeV we
will have to see many more of energy higher (even 1000 EeV). We should point
out that this result is independent of the mass of the monopole, since the
kinetic energy acquired by the monopole does not depend on the mass. We can see
that monopoles accelerate at much higher energies than those quoted by
\cite{Weiler_1,Weiler_2} and with an spectrum which is essentially flat over a
large energy region ($\sim$ one order of magnitude). This spectrum is clearly
not compatible with the observed spectrum, which falls off steeply with energy
(slope $\sim$ -2.7). The result is independent of the assumed magnetic field
model and it is true so long as the random component is smaller than the
regular component.

\begin{figure}
\epsfxsize=10cm
\begin{center}
\mbox{\epsfig{file=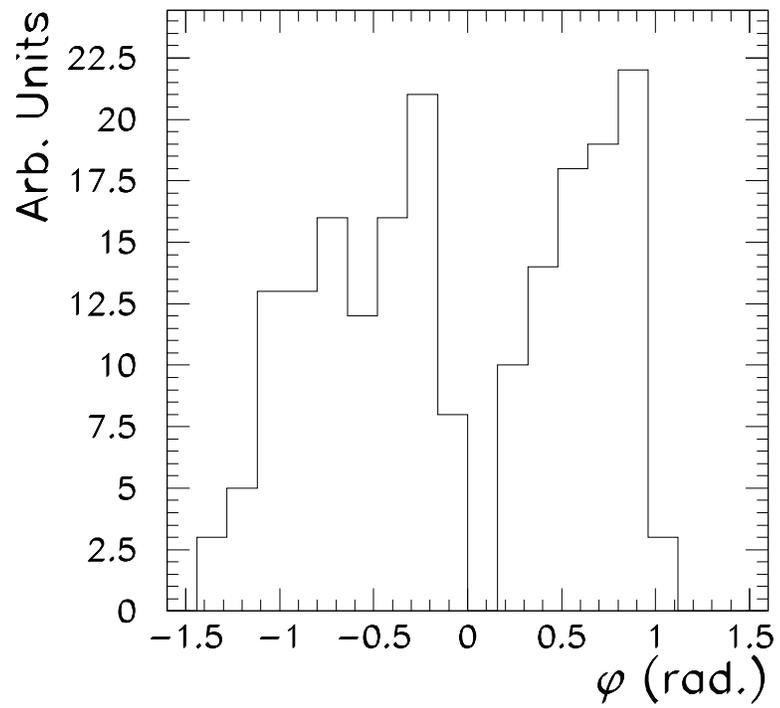}}
\end{center}
\caption{Arrival direction (in galactic longitude, radians) for monopoles
coming from the center of the galaxy.}
\label{arrival_square}
\end{figure}

The arrival direction of monopoles in this scenario is plotted in
fig.\ref{arrival_square} where we plot the $\phi$ angle distribution
($\phi=0^\circ$ corresponds to the galactic center) for the model A. We can see
that there are ''forbidden'' directions: monopoles don't come directly from the
galactic center. Also they don't come at $\phi= 180^\circ$. For both models the
galactic magnetic field at the position of the Earth is directed almost in the
$\phi = 90^\circ$ direction (with a small pitch angle) which aligns the
monopoles irrespectively of their initial direction.

One may argue that there is no reason why monopoles should be distributed near
the center of the galaxy. Due to their low mass, these monopoles are always
relativistic and therefore they are not confined by the gravitational galactic
field which is in fact negligible for them. And probably they are not confined
either in supernovas. We can see this easily. A monopole is accelerated by a
magnetic field to:
\begin{equation}
\gamma-1 \sim \frac{q_M \; B L}{M} \sim 0.02 \; \frac{B}{3 10^{-6} G} \; 
\frac{L}{1 pc}.
\end{equation}
In other words, in only 1 pc the monopole could accelerate to a velocity much
above the escape velocity for typical galactic magnetic fields. This also
justifies a posteriori our neglecting of the gravitational field.

\subsection{Scenario B}

Let's assume next that monopoles are accelerated somehow in the extragalactic
magnetic field and that they penetrate through the galaxy and arrive to Earth.
We simulate this by back propagating the monopoles from the Earth with an
energy $E_0$ back in time until a given distance is reached. We find that for
almost all arrival directions the monopoles, in fact, have to loose energy in
order to reach the Earth. This is a consequence of the ''spiral'' form of the
magnetic field: Only a small cone of angles are in favour of the magnetic
field, and this is only for a (relatively) small distance, for the magnetic
field changes direction and then the monopole finds itself running ''against
the wind''. This is independent of the sign of the magnetic charge of the
monopole and also of the model chosen for the magnetic field.

\begin{figure}
\epsfxsize=10cm
\begin{center}
\mbox{\epsfig{file=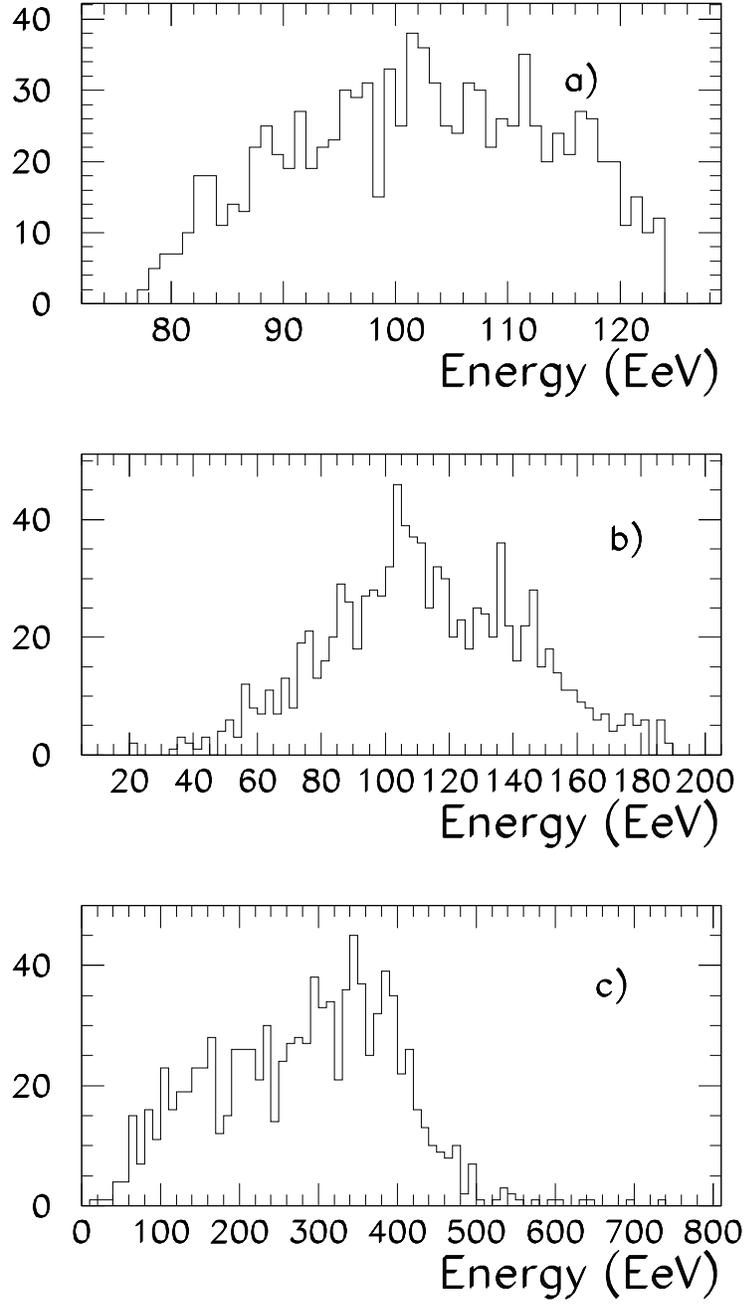}}
\end{center}
\caption{Energy spectrum of monopoles reaching the Earth with fix energy,
$E= 100$ EeV, after they are backpropagated a) 300 pc, b) 900 pc, c) 3 kpc.}
\label{cube}
\end{figure}

Then for most of the arrival directions we find that monopoles are in fact
deaccelerated in the galactic magnetic field and that in order to arrive at
Earth with, say, $\sim $ 100 EeV they needed to have much bigger energies
initially. This can be seen in fig.\ref{cube} where we plot the energy
distribution for monopoles arriving at Earth with energy 100 EeV and back
propagated to a distance of 300 pc (a), 900 pc (b) and 3 kpc (c). At a distance 
of 3 kpc only a small fraction are really accelerated (the average energy to
reach the Earth with 100 EeV is 270 EeV).

\section{Comparison with observed HECR}

What about the observed high energy events? There are 8 observed events above
100 EeV. All, but one, arrive in a direction almost perpendicular to the local
magnetic field which implies that irrespectively of being a monopole or an
antimonopole these directions are not favoured by the local magnetic field.
Only the Akeno event (pair event n.3 in ref.\cite{Hayashida_1}) is in a
direction favoured by the local magnetic field, if this event was an
antimonopole. This must be necessarily an odd and very improbable situation
because one should expect monopoles to come aligned with the local magnetic
field. In fact observation suggest that the arrival direction of the highest
energy events are correlated with the supergalactic plane \cite{Stanev_2} which
for the observed region of the supergalactic plane is almost perpendicular to
the local magnetic field.

Events with energy lower than around 100 EeV are not assumed to be monopoles
since they would not have energy enough to initiate showers. But the Stanev et
al. \cite{Stanev_2} study (and see also the Akeno study of their events
\cite{Hayashida_1}) suggest that the correlation is valid for lower energy
events too ($E > 40$ EeV), hinting towards the same origin for both types of
events. In addition many of the highest energy events are paired with other of
lower energy. This is true for the Fly's Eye event (which is paired with one
Akeno event and with the Yakutsk event) and for the second's world highest
energy event. These correlations can be easily accounted for in a conventional
scenario whereas in the monopole case they are difficult to explain.

\begin{figure}
\epsfxsize=10cm
\begin{center}
\mbox{\epsfig{file=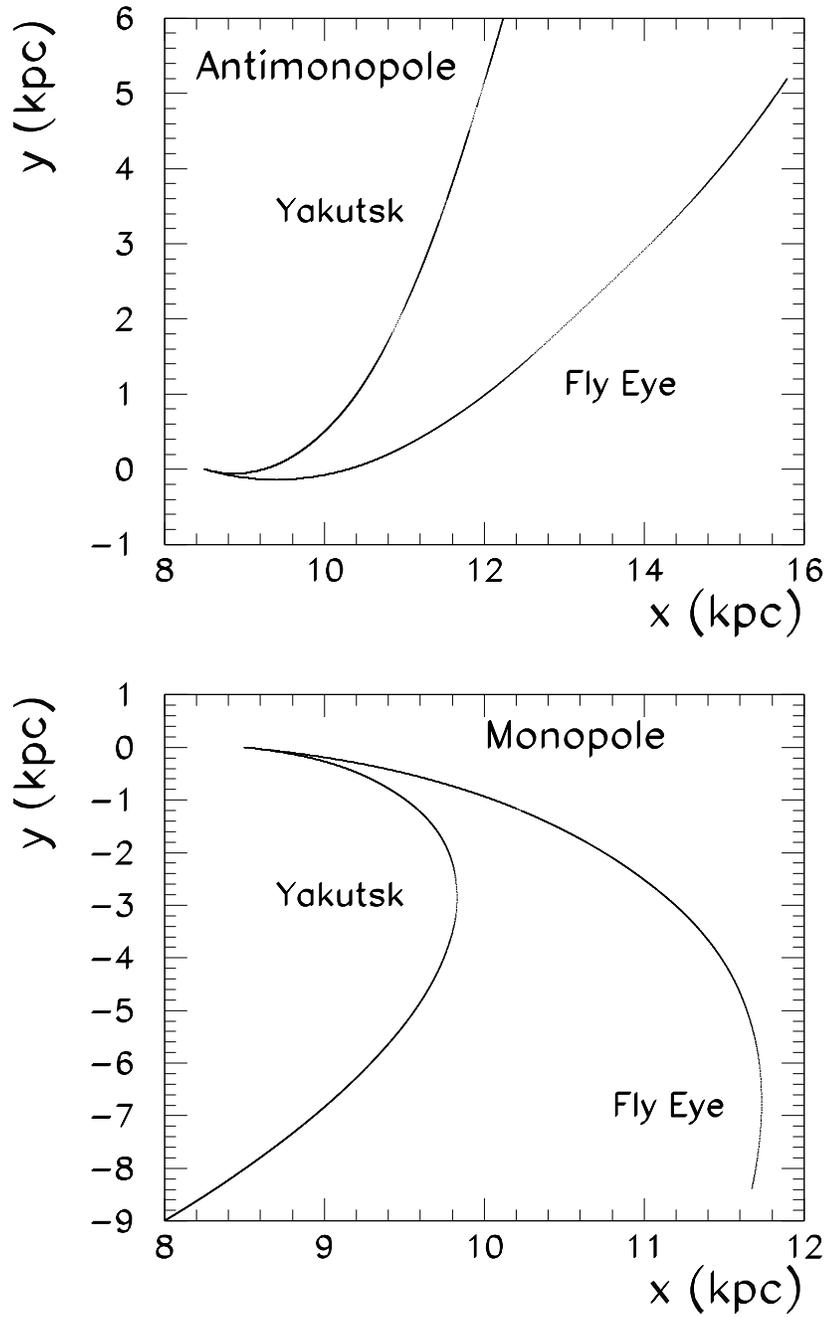}}
\end{center}
\caption{Trajectory on the $x y$ plane (in kpc) of the highest energy events
detected by Fly's Eye and by Yakutsk, assuming they are antimonopoles (upper
figure) or monopoles (lower figure).}
\label{fly1}
\end{figure}

\begin{figure}
\epsfxsize=10cm
\begin{center}
\mbox{\epsfig{file=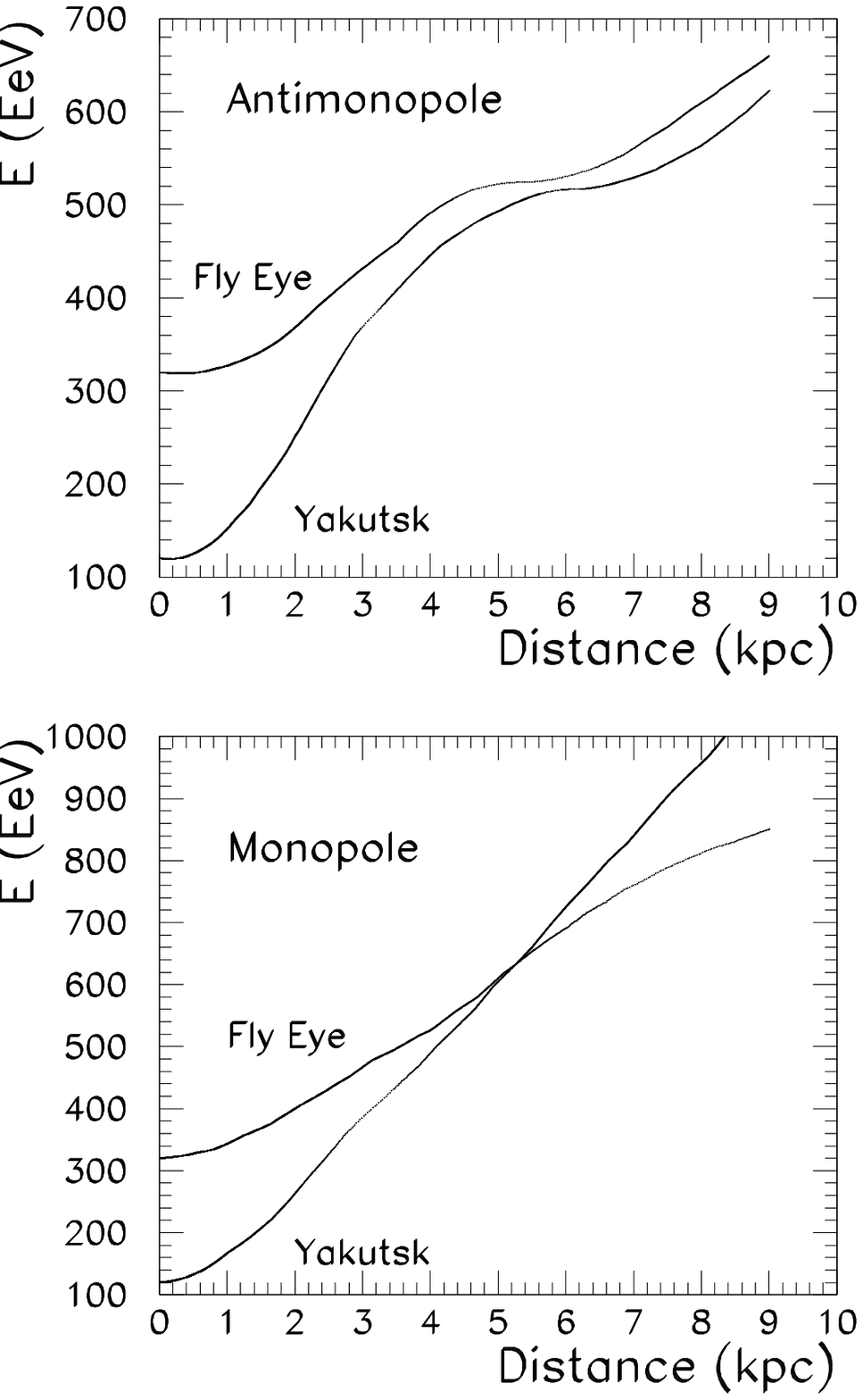}}
\end{center}
\caption{Monopole energy as a function of the distance to the Earth for the two
events of the previous figure.}
\label{fly2}
\end{figure}

Let's consider the Fly's Eye event. It has the same arrival direction (within
errors) as the highest event detected by Yakutsk, but it has a higher energy.
It is obvious that if we back propagate them from the Earth to a given
distance, the original position for both events must be very different (and
also the original momentum). So we have the strange situation of two monopoles
coming from very different points of the galaxy but they manage to arrive at
the same point with the same direction. As mentioned before, this direction is
not favoured by the magnetic field. We can see in fig.\ref{fly1} the trajectory
for both events in the $x y$ plane and in fig.\ref{fly2} the energy as a
function of the distance to the Earth (for both a monopole and an antimonopole
origin). In any case, the particle must be deaccelerated by the magnetic field.
Certainly it must be very improbable such situation but giving an actual number
for the probability is difficult.

\begin{figure}
\epsfxsize=10cm
\begin{center}
\mbox{\epsfig{file=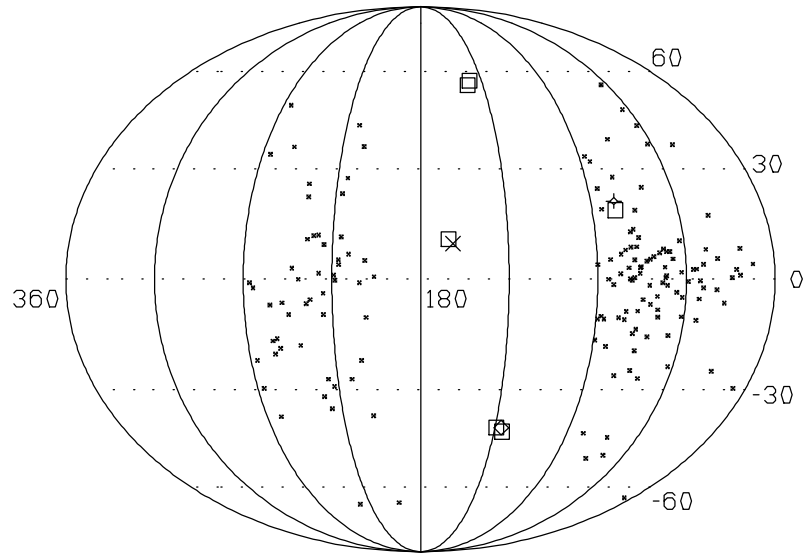}}
\end{center}
\caption{Arrival direction in galactic coordinates of monopoles seed uniformly
in the galaxy and arriving near the Earth (small dots). High energy observed
events are represented by squares (Akeno events), crosses (Fly's Eye) and
diamonds (Haverah Park).}
\label{flux}
\end{figure}

In order to give a partial answer, we ran the following simulation. Monopoles
are seeded at a random position in a cube of 20 kpc centered in the Galaxy 
with an initial random energy which is assumed to be a gaussian of mean energy
100 EeV (this is justified by the previous simulations). They are propagated in
the galactic field until i) the distance to the Earth is bigger than 20 kpc
where they are assumed to go out of the galaxy and are dropped or ii) the
distance to the Earth is less than 4 kpc and then they are kept. The flux at
the Earth is assumed to be same as the flux in the sphere of 4 kpc around the
Earth. The final distance, 4 kpc, is a compromise between computer time and
getting the closest possible to the Earth, but it is a conservative assumption.
Getting closer to the Earth can only collimate more the fluxes and therefore
give more acute differences with the observed flux. We present in
fig.\ref{flux} the arrival direction in galactic coordinates of the monopoles
which satisfy the above criteria for the model A. For the sake of comparison we
also show the arrival direction of the most energetic cosmic rays which were
available to us. As said previously we observe that no monopole arrived in the
direction of the observed events, all of them arrive with either $\sim
90^\circ$ or $\sim 270^\circ$ longitude, as expected, mainly polarized
according to the magnetic lines. Model B gives qualitatively the same results.
We see that arrival directions do not cluster around any particular point. 

We may conclude that in order to observe one high energy event, for instance,
in the direction of the Fly's Eye event we should have observed about 200 in
the directions of the galactic magnetic field with similar (or higher) energy.
This is just a rough estimation due to the approximations involved in the box
chosen and since we don't take into account the exposure of the detectors to
the different regions of the galactic plane but it is a significative one.
Alternatively, if we normalize the calculated angle integrated flux to the
observed flux, we see that the flux in the arrival direction of the observed
events must be $\sim$ 100 lower than observed and no event could have been
seen. 

\section{Conclusions}

In conclusion, magnetic monopoles are very unlikely the cosmic ray particles at
the highest energies. Expected fluxes from monopoles are highly anisotropic,
pointing towards the magnetic lines near the Earth. Moreover, the energy of the
monopoles should be much higher than observed and the spectrum, basically flat,
contrary to observation. These results are based only in the structure of the
magnetic field. A random component of the magnetic field could change this
picture only if it is larger than the regular component.

\centerline{Acknoledgements}

The authors thank M.C. Gonzalez-Garcia for discussions on the subject. RAV
thanks the IFT for its kind hospitality and for providing us with computer
resources. RAV's work was supported by FAPESP.



\end{document}